\documentstyle[12pt]{article}
\def\b{\beta}
\def\t{\theta}

\def\ch{ch}
\def\sh{sh}

\def\cosh{cosh}
\def\ve{\varepsilon}
\def\non{\nonumber}

\begin{document}

\newpage
\pagestyle{empty}
\begin{flushright}
DTP/95-11 \\
hep-th/9504018 \\
April 1995
\end{flushright}
\begin{center}
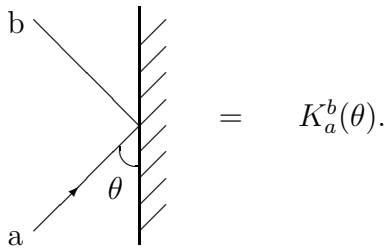
{\Large {\bf Boundary Reflection Matrix \\
in Perturbative Quantum Field Theory}}
\end{center}
\begin{center}
{\bf J. D. Kim\footnote{jideog.kim@durham.ac.uk} \footnote{On leave
of absence from Korea Advanced Institute of Science and Technology} }  \\
{\it Department of Mathematical Sciences, \\
University of Durham, Durham DH1 3LE, U.K. }  \\
\end{center}
\vspace{3cm}

\begin{center}
ABSTRACT
\end{center}

We study boundary reflection matrix
for the quantum field theory defined on a half line
using Feynman's perturbation theory. The boundary reflection
matrix can be extracted directly from the two-point correlation function.
This enables us to determine the boundary reflection matrix
for affine Toda field theory with the Neumann boundary condition modulo
\lq a mysterious factor half'.

\newpage
\pagestyle{plain}

\section*{I. Introduction}
The $S$-matrix of (massive) integrable quantum field theory in 1+1 dimensions
can be studied by several different methods.
The high brow technology to construct $S$-matrix
is based on the symmetry principle
such as Yang-Baxter equation, unitarity, crossing relation,
real analyticity and bootstrap equation\cite{ZZ,BCDS,CM}.
This program entirely relies on the assumed quantum integrability of the model
and produces an $S$-matrix which is exact up to all loop order.

Despite its beautiful nature, this method inevitably needs
additional information.
Furthermore, there is an inherent so-called CDD ambiguity.
To back up the situation, Feynman's perturbation theory
has been used and shown to agree well with the conjectured \lq minimal'
$S$-matrices\cite{BCDS,CM,AFZ,BCDS2,BS,CKK,BCKKR,SZ}.
This may also be considered
as the strong evidence for the assumed quantum integrability.
In perturbation theory, $S$-matrix is extracted
from the four-point correlation function with LSZ reduction formalism.

About a decade ago, integrable quantum field theory on a half line
$(-\infty < x \leq 0)$
was studied using symmetry principles under the assumption that
the integrability of the model remains intact\cite{Che}.
The boundary Yang-Baxter equation, unitarity relation
for boundary reflection matrix $K_a^b(\t)$
which is conceived to describe the scattering process off a wall
was introduced\cite{Che}.

\linethickness{0.5pt}
\begin{picture}(350,130)(-70,-25)
\put(100,95){\line(0,-1){90}}
\put(100,50){\line(-1,1){40}}
\put(100,50){\line(-1,-1){40}}
\put(75,25){\vector(1,1){2}}
\put(50,85){b}
\put(50,5){a}
\put(100,42){\oval(15,15)[b,l]}
\put(88,22){$\t$}
\put(100,80){\line(1,1){10}}
\put(100,70){\line(1,1){10}}
\put(100,60){\line(1,1){10}}
\put(100,50){\line(1,1){10}}
\put(100,40){\line(1,1){10}}
\put(100,30){\line(1,1){10}}
\put(100,20){\line(1,1){10}}
\put(100,10){\line(1,1){10}}
\put(130,50){=}
\put(160,50){$K_a^b(\t)$.}
\put(20,-20){Figure 1. Boundary Reflection Matrix.}
\end{picture}

Recently, boundary crossing relation was introduced\cite{GZ}.
In fact, the boundary crossing relation is
automatically satisfied if the boundary bootstrap equation is
satisfied\cite{Sasaki}.
Subsequently, some exact boundary reflection matrices
have been constructed\cite{GZ,Gh,Sasaki,FK,CDRS} for affine Toda field
theory(ATFT).
However, it turns out that there is a plethora of solutions for
boundary reflection matrix
despite requiring the \lq minimality' assumption
which has been effective in the $S$-matrix theory on a full line.
Furthermore, it is unknown what
particular boundary condition(or boundary potential) actually
corresponds to a particular solution.

In order to have a direct access to the boundary reflection matrix,
it seems compelling to study the boundary system from the
Lagrangian quantum field theory.
In fact, several studies on the boundary system
has already been done\cite{Sym,DD,BM}
in the Lagrangian quantum field theory context.
However, the boundary reflection matrix has never been discussed in
this framework, yet.

On the other hand, the ordinary LSZ reduction method which does the job
to extract the $S$-matrix from the off-shell correlation functions
becomes inapplicable for the quantum field theory on a half line,
since the momentum eigenstates with a definite sign in the asymptotic region
do not satisfy the boundary condition at the origin in space.

In this paper, we propose a method to extract boundary reflection matrix
directly from the two-point correlation function.
In section II, we describe the formalism.
In section III, we present the one loop result for the sinh-Gordon model
(or $a_1^{(1)}$ affine Toda theory).
In section IV, we present the one loop result for the Bullough-Dodd model
(or $a_2^{(2)}$ affine Toda theory).
We also give a conjecture for the exact boundary reflection matrix
guided from this one loop result.
This model fully utilises all possible
Feynman diagrams because it has three point self-coupling as well as
four-point coupling.
Finally, we give some discussions in section V.

\section*{II. Boundary Reflection Matrix}
We are mainly concerned with affine Toda field theory with integrable
boundary interaction
though the formalism may be applicable for any quantum field theory.

To begin with, we review the
two-point function for the model on a full line.
The  bosonic ATFT\cite{BCDS} is defined by the following Lagrangian
density based on a Lie algebra $g$ with rank $r$.
\begin{equation}
{\cal{L}}(\Phi) = \frac{1}{2}\partial_{\mu}\phi^{a}\partial^{\mu}\phi^{a}
-\frac{m^{2}}{\b^{2}}\sum_{i=0}^{r}n_{i}e^{\b \alpha_{i} \cdot \Phi},
\end{equation}
where
\begin{displaymath}
\alpha_{0} = -\sum_{i=1}^{r}n_{i}\alpha_{i},~~ \mbox{and}~~ n_{0} = 1 .
\end{displaymath}
The field $\phi^{a}$ ($a=1,\cdots,r$) is $a$-th component of the scalar field
$\Phi$,
and $\alpha_{i}$ ($i=1,\cdots,r$) are simple roots of $g$ normalized
so that the universal function $B(\b)$
through which the dimensionless coupling
constant $\b$ appears in the $S$-matrix takes the following form:
\begin{equation}
B(\b)=\frac{1}{2\pi}\frac{\b^2}{(1+\b^2/4\pi)}.
\label{Bfunction}
\end{equation}
The $m$ sets the mass scale and the $n_i$s are the so-called Kac labels
which are characteristic integers defined for each Lie algebra.

The two-point function at tree level is given by the Feynman propagator:
\begin{equation}
G (t',x';t,x)=\int \frac{d^2 p}{(2 \pi)^2} \frac{i}{p^2-m_a^2+i \ve}
  e^{-i w (t'-t)+i k (x'-x)},
\label{FGF}
\end{equation}
where $p=(w,k)$ is the two dimensional energy-momentum and
$m_a$ is the mass of the particle in the original Lagrangian.
As is well known, this two-point function depends
only on the difference of its arguments
and accommodates contributions coming from the positive energy states
as well as the negative energy states
depending on the sign of the difference of the time arguements.
This physical feature is usually implemented by $i \ve$ prescription
or choice of $w$-contour.

At one loop order, there are three types of Feynman diagram contributing to
the two-point correlation function as depicted in Figure 2.

\linethickness{0.5pt}
\begin{picture}(350,175)(-50,-110)
\put(0,0){\circle{40}}
\put(-20,50){\line(0,-1){100}}
\put(-15,35){a}
\put(-15,-35){a}
\put(25,0){b}
\put(-18,-70){Type I.}
\put(170,0){\circle{40}}
\put(130,0){\line(1,0){20}}
\put(130,50){\line(0,-1){100}}
\put(135,35){a}
\put(135,-35){a}
\put(137,5){b}
\put(195,0){c}
\put(135,-70){Type II.}
\put(300,0){\circle{40}}
\put(300,20){\line(0,1){30}}
\put(300,-20){\line(0,-1){30}}
\put(305,35){a}
\put(305,-35){a}
\put(270,0){b}
\put(325,0){c}
\put(280,-70){Type III.}
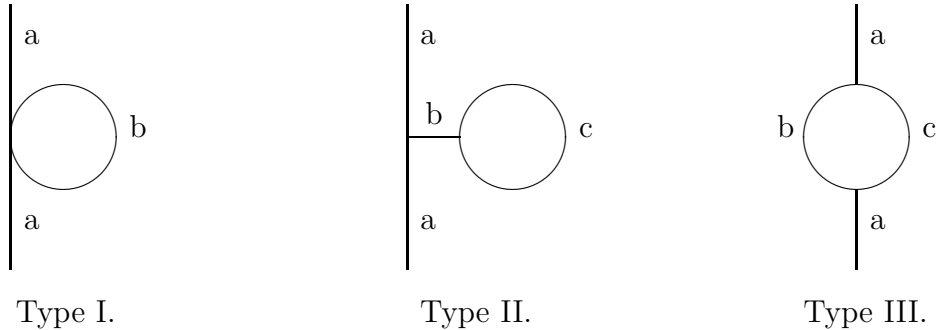
\put(10,-100){Figure 2. Diagrams for the one loop two-point function.}
\end{picture}

Type I, II diagrams have logarithmic infinity independent of the
external energy-momenta and are the only divergent diagrams in 1+1 dimensions.
This infinity is usually absorbed into
the infinite mass renormalization.
Type III diagrams have finite corrections
depending on the external energy-momenta
and produces a double pole to the two-point correlation function.

The remedy for these double poles is to introduce a counter term
to the original Lagrangian to cancel this term(or to renormalize the mass).
In addition, to maintain the residue of the pole, we have to
introduce wave function renormalization.
Then the renormalized two-point correlation function remains the same
as the tree level one with renormalized mass $m_a$,
whose ratios are the same as the classical value.
This mass renormalization procedure can be generalized
to arbitrary order of loops.

Now we consider the model on a half line($-\infty < x \leq 0$).
The action is defined as follows,
\begin{equation}
S(\Phi) = \int_{-\infty}^{0} dx \int_{-\infty}^{\infty} dt
\left ( \frac{1}{2}\partial_{\mu}\phi^{a}\partial^{\mu}\phi^{a}
-\frac{m^{2}}{\b^{2}}\sum_{i=0}^{r}n_{i}e^{\b \alpha_{i} \cdot \Phi}
\right ),
\end{equation}
The above simple action may have additional boundary potential
maintaining the integrability.
Non-trivial boundary potentials which do not destroy the integrability
have been determined at the classical level\cite{GZ,CDRS,CDR,Mac,BCDR}.
The stability of the model with boundary potential
has also been discussed\cite{CDR,FR}.
Here we consider the model with no boundary potential,
which corresponds to the Neumann boundary condition:
$\frac{ \partial \phi^a} {\partial x} =0$ at $x=0$.
This case is believed to be quantum stable in the sense that
the existence of a boundary does not change
the structure of the spectrum.

At tree level, two-point correlation functions are given by a sum of
a direct contribution and a reflected one which may be considered as coming
from
the image point,
\begin{eqnarray}
G_N (t',x';t,x) &=& G(t',x';t,x) + G(t',x';t,-x) \\
             &=& \int \frac{d^2 p}{(2 \pi)^2} \frac{i}{p^2-m_a^2+i \ve}
  e^{-i w (t'-t)}
 (  e^{i k (x'-x)} +  e^{i k (x'+x)} ).  \non
\end{eqnarray}
We may use the $k$-integrated version.
\begin{equation}
G_N (t',x';t,x) = \int \frac{d w}{2 \pi} \frac{1}{2 \bar{k}}
  e^{-i w (t'-t)} (  e^{ i \bar{k} |x'-x| } +  e^{-i \bar{k} (x'+x)} ),
{}~~ \bar{k}=\sqrt{w^2-m_a^2}.
\end{equation}
We find that the unintegrated version is very useful to extract the
asymptotic part of the two-point correlation function far away from the
boundary.

To compute two-point correlation functions at one loop order,
we follow the idea of the conventional perturbation theory\cite{Sym,DD,BM}.
That is, we generate the relevant Feynman diagrams
and then evaluate each of them by using the zero-th order
two-point function for each line occurring in the Feynman diagrams.

Type I diagram gives the following contribution.
\begin{equation}
\int_{-\infty}^{0} d x_1  \int_{-\infty}^{\infty} d t_1
 G_N (t,x;t_1,x_1) ~ G_N (t',x';t_1,x_1) ~ G_N (t_1,x_1;t_1,x_1).
\label{TypeI}
\end{equation}
Let us take a close look at $ G_N (t_1,x_1;t_1,x_1) $.
\begin{eqnarray}
G_N (t_1,x_1;t_1,x_1) &=& \int \frac{d^2 p_1}{(2 \pi)^2} \frac{i}{p_1^2-m_b^2+i
\ve}
  ( 1 +  e^{i k_1 2 x_1} ).
\end{eqnarray}
The first term is the ordinary infinite mass renormalization term
as for the full line theory.
To cancel this, we introduce a counter term exactly the same as for the full
line.
We should not simply discard the second term and this term contributes to
the boundary reflection matrix.
We evaluate $t_1$ integral in Eq.(\ref{TypeI}),
giving energy conservation at the interaction vertex.
The $x_1$ integral gives \lq spatial momentum conservation' as follows:
\begin{equation}
k ~\pm k' +2 k_1 =0.
\label{SMCI}
\end{equation}

After the integrations of loop variables and energy conserving delta function
which resulted from $t_1$ integral,
we get the following result from Type I diagram,
\begin{equation}
\int \frac{dw}{2 \pi} \frac{dk}{2 \pi} \frac{dk'}{2 \pi}
  ~~ e^{-iw(t'-t)}  e^{i (kx+k'x')}
 \frac{i}{w^2-k^2-m_a^2+i \ve} \frac{i}{w^2-k'^2-m_a^2+i \ve} ~~ I_1,
\end{equation}
\begin{displaymath}
I_1 \equiv \frac{1}{2 \sqrt{k_1^2+m_b^2} },
\end{displaymath}
where $k_1$ is defined in Eq.(\ref{SMCI}).

{}From Type II diagram, we can read off the following expression:
\begin{equation}
\int_{-\infty}^{0} d x_1 d x_2 \int_{-\infty}^{\infty} d t_1 d t_2
 G_N (t,x;t_1,x_1) ~ G_N (t',x';t_1,x_1) ~ G_N (t_1,x_1;t_2,x_2)
\end{equation}
\begin{displaymath}
{}~~~~~~~~~~ G_N (t_2,x_2;t_2,x_2).
\end{displaymath}
Similarly as for the Type I diagram, $ G_N (t_2,x_2;t_2,x_2) $ contains
the ordinary infinite tadpole term. By introducing infinite mass
renormalization, we can discard this tadpole term.

The $t_1$ and $t_2$ integral give energy conservations at each vertex.
The $x_1$ and $x_2$ integral gives \lq spatial momentum conservation' as
follows,
\begin{eqnarray}
k ~\pm k' + k_1 =0, &  k_1 + 2 k_2 =0.
\label{SMCII}
\end{eqnarray}

After the integrations of loop variable $w_2$, momentum conserving
delta functions as in Eq.(\ref{SMCII}) and energy conserving
delta functions $\delta(w_1), \delta(w'+w)$, we get the following
result from Type II diagram,
\begin{equation}
\int \frac{dw}{2 \pi} \frac{dk}{2 \pi} \frac{dk'}{2 \pi}
  ~~ e^{-iw(t'-t)}  e^{i (kx+k'x')}
 \frac{i}{w^2-k^2-m_a^2+i \ve} \frac{i}{w^2-k'^2-m_a^2+i \ve} ~~ I_2,
\end{equation}
\begin{displaymath}
I_2 \equiv \frac{-i}{k_1^2+m_b^2} \frac{1}{2 \sqrt{k_2^2+m_c^2} }.
\end{displaymath}

Type III diagram gives following contribution:
\begin{equation}
\int_{-\infty}^{0} d x_1  d x_2 \int_{-\infty}^{\infty} d t_1 d t_2
 G_N (t,x;t_1,x_1) ~ G_N (t',x';t_2,x_2) ~ G_N (t_2,x_2;t_1,x_1)
\end{equation}
\begin{displaymath}
G_N (t_2,x_2;t_1,x_1).
\end{displaymath}

The $t_1$ and $t_2$ integral give energy conservations $\delta(w+w_1+w_2),
\delta(w_1+w_2-w')$ at each vertex.
The $x_1$ and $x_2$ integral give \lq spatial momentum conservations' as
follows,
\begin{eqnarray}
 k ~\pm k_1 \pm k_2 =0, & \pm k_1 \pm k_2 + k'=0.
\label{SMCIII}
\end{eqnarray}
Among the 16 possible combinations of signs in front of each spatial momentum
in the above equation, the same 8 combinations as the momenta
conservation on a full line gives exactly the same finite mass renormalization.
The other combinations give the following result from Type III diagram,
\begin{equation}
\int \frac{dw}{2 \pi} \frac{dk}{2 \pi} \frac{dk'}{2 \pi}
  ~~  e^{-iw(t'-t)}  e^{i (kx+k'x')}
   \frac{i}{w^2-k^2-m_a^2+i \ve} \frac{i}{w^2-k'^2-m_a^2+i \ve} ~~ I_3,
\label{I-III}
\end{equation}
\begin{displaymath}
I_3 \equiv
  \frac{1}{4}
  ( \frac{i}{2 \bar{w}_1 (\bar{w}_1-\tilde{w}_1^+) (\bar{w}_1-\tilde{w}_1^-)}
  + \frac{i}{(\tilde{w}_1^+ -\bar{w}_1) (\tilde{w}_1^+ +\bar{w}_1)
        (\tilde{w}_1^+ -\tilde{w}_1^- )  } ),
\end{displaymath}
where we included $\frac{1}{4}$ which was introduced
while we were extending the domain of $x_1, x_2$ integrations to a full line
to allow the delta function and we introduced the following notations.
\begin{eqnarray}
\bar{w}_1=\sqrt{k_1^2+m_b^2}, &
\tilde{w}_1^+ =w+\sqrt{k_2^2+m_c^2}, & \tilde{w}_1^- =w-\sqrt{k_2^2+m_c^2}.
\end{eqnarray}
It should be remarked that this term should be symmetrized with respect
to $m_b, m_c$.

Now we propose a method to extract boundary reflection matrix
directly from the two-point correlation function.
The general form of each contributions
coming from type I,II and III diagrams can be written as follows:
\begin{equation}
\int \frac{dw}{2 \pi} \frac{dk}{2 \pi} \frac{dk'}{2 \pi}
   e^{-iw(t'-t)}  e^{i (kx+k'x')}
 \frac{i}{w^2-k^2-m_a^2+i \ve} \frac{i}{w^2-k'^2-m_a^2+i \ve}
 I(w,k,k').
\label{General}
\end{equation}
Contrary to the other terms which resemble those of a full line,
this integral has two spatial momentum integration.
First, let us consider the $k'$ integration.
There are two contributions.
One comes from the usual pole contribution of the propagator
and the other one from the poles and the branch cuts of $I$ function if any.
For the $k$ integration, the similar consideration can be done.
Here we simply state that the contributions
other than the usual pole contributions
coming from each poles of the external propagators turn out
to be exponentially damped as $x, x'$ go to $-\infty$.

In this way, we can get a method to compute
elastic boundary reflection matrix $K_a(\t)$ defined
as the coefficient of the reflected term of the exact two-point correlation
function in the asymptotic region far away from the boundary.
\begin{displaymath}
\int \frac{dw}{2 \pi} e^{-iw(t'-t)} \frac{1}{2 \bar{k}}
 (  e^{i \bar{k} |x'-x|} +K_a(w)  e^{-i \bar{k} (x'+x)} ),
  ~~ \bar{k}=\sqrt{w^2-m_a^2}.
\end{displaymath}
$K_a(\t)$ is obtained using $w=m_a \cosh\t$.

Here we list each one loop contribution to $K_a(\t)$
from the three types of diagram depicted in Figure 2:
\begin{eqnarray}
K_a^{(I)}(\t) &=& \frac{1}{2 m_a \sh\t} ( \frac{1}{2 \sqrt{m_a^2
\sh^2\t+m_b^2}}
    +\frac{1}{2 m_b} ) ~C_1 ~S_1,  \\
K_a^{(II)}(\t) &=& \frac{1}{2 m_a \sh\t}
  ( \frac{-i}{ (4 m_a^2 \sh^2\t +m_b^2) 2 \sqrt{m_a^2 \sh^2\t+m_c^2}}
    +\frac{-i}{ 2 m_b^2 m_c} ) ~C_2 ~S_2,  \\
K_a^{(III)}(\t) &=& \frac{1}{2 m_a \sh\t}
  ( 4 I_3(k_1=0,k_2=k)+4 I_3(k_1=k,k_2=0) ) ~ C_3 ~S_3,
\label{K-III}
\end{eqnarray}
where $I_3$ is the function defined in Eq.(\ref{I-III}),
and the factor 4 in front of it accounts for
the fact that there are four combinations in Eq.(\ref{SMCIII}) which
give the same result.
The $C_i, S_i$ denote numerical coupling factors and symmetry factors,
respectively.

\section*{III. Example I : $a_1^{(1)}$ affine Toda theory}
For the sinh-Gordon model, only Type I diagram in Figure 2 contributes to
one loop two-point correlation function since there is no three-point
self coupling.
We have to fix the normalization of roots so
that the standard $B(\b)$ function takes the form given in
Eq.(\ref{Bfunction}).

We use the Lagrangian density given as follows.
\begin{eqnarray}
{\cal{L}}(\phi) &=& \frac{1}{2}\partial_{\mu}\phi \partial^{\mu}\phi
-V(\phi), \\
 V(\phi) &=& \frac{m^{2}}{4 \b^{2}}
                ( e^{\sqrt{2} \b \phi} + e^{-\sqrt{2} \b \phi} -2) \\
         &=& \frac{1}{2}m^2 \phi^2-\frac{1}{12} m^2 \b^2 \phi^4 +O(\b^4).  \non
\end{eqnarray}

The scattering matrix for the elementary scalar of this model is\cite{ZZ}
\begin{equation}
S(\t)=\frac{ (0) (2) }{ (B) (2-B) }.
\end{equation}
Here $B$ is the same function defined in Eq.(\ref{Bfunction}) and
we used the usual notation of building block\cite{BCDS} as follows.
\begin{equation}
 (x) = \frac{ \sh( \t /2 + i \pi x /2 h )}
            { \sh( \t /2 - i \pi x /2 h )}.
\label{Blockx}
\end{equation}
For the sinh-Gordon model, $h=2$ and from now on we set $m=1$.

The result coming from Type I diagram is
\begin{equation}
K(\t) = \frac{1}{2 \sh\t} ( \frac{1}{2 \ch\t}+\frac{1}{2})
    \times (\frac{-i}{12} \b^2) \times 12.
\end{equation}
It turns out that this is too large by a factor 2
to satisfy the crossing unitarity relation at one loop order.
\begin{equation}
K(\t) ~ K(\t-i \pi) = S(2 \t).
\end{equation}

So we need to include an extra factor $\frac{1}{2}$ into our formulae
in Eq.(\ref{K-III}), although we do not understand the reason.
Then, we find that the formulae in Eq.(\ref{K-III}) with the extra factor
$\frac{1}{2}$ work for any theory.

On the other hand, there are two \lq minimal' boundary reflection matrices
which are meromorphic in terms of rapidity variable
are known for $a_1^{(1)}$ model\cite{Sasaki,FK}.
One of them agrees with the perturbative result.
\begin{equation}
K(\t)=[ 1 /2 ],
\end{equation}
where
\begin{equation}
 [ x ] = \frac{ (x-1/2) (x+1/2)} {(x-1/2+B/2) (x+1/2-B/2)}.
\end{equation}

\section*{IV. Example II : $a_2^{(2)}$ affine Toda theory}
The Bullough-Dodd model has three-point self coupling as well as four-point
coupling.
So all possible three types of diagram in Figure 2 contribute to one loop
two-point correlation function. We have to fix the normalization of roots so
that the standard $B(\b)$ function takes the form given in
Eq.(\ref{Bfunction}).

We use the Lagrangian density given as follows.
\begin{eqnarray}
{\cal{L}}(\phi) &=& \frac{1}{2}\partial_{\mu}\phi \partial^{\mu}\phi
-V(\phi), \\
 V(\phi) &=& \frac{m^{2}}{6 \b^{2}} (2 e^{\b \phi} + e^{-2 \b \phi} -3) \\
         &=& \frac{1}{2}m^2 \phi^2-\frac{1}{6} m^2 \b \phi^3
             +\frac{1}{8} m^2 \b^2 \phi^4 +O(\b^3).  \non
\end{eqnarray}
The scattering matrix for the elementary scalar of this model is\cite{AFZ}
\begin{equation}
S(\t)=\frac{ (0) (2) (1) (3) }{ (B) (2-B) (1+B) (3-B) }.
\end{equation}
For the Bullough-Dodd model, $h=3$.

The result coming from Type I diagram is
\begin{equation}
K^{(I)}= \frac{1}{4 \sh\t} ( \frac{1}{2 \ch\t}+\frac{1}{2})
    \times (\frac{-i}{8} \b^2) \times 12.
\end{equation}
The result coming from Type II diagram is
\begin{equation}
K^{(II)}= \frac{1}{4 \sh\t}  (\frac{1}{(4 \sh^2\t +1)}\frac{-i}{2
\ch\t}-\frac{i}{2})
 \times (\frac{i}{6} \b)^2 \times 18.
\end{equation}
For type III diagram, when $ k_1=0, k_2=k$,
\begin{equation}
\bar{w}_1=1,      ~~ \tilde{w}_1^+ = 2 \ch\t, ~~ \tilde{w}_1^- =0,
\end{equation}
and when $k_1=k, k_2=0$,
\begin{equation}
\bar{w}_1=\ch\t, ~~ \tilde{w}_1^+ = \ch\t+1 , ~~ \tilde{w}_1^- =\ch\t-1.
\end{equation}
The result coming from Type III diagram is
\begin{equation}
K^{(III)}= \frac{1}{4 \sh\t} (\frac{i}{2(1-2 \ch\t)}
   + \frac{i}{(2 \ch\t-1)(2 \ch\t+1) 2 \ch\t} + \frac{i}{-2 \ch\t}
       + \frac{i}{(2 \ch\t+1) 2})
\end{equation}
\begin{displaymath}
 \times (\frac{i}{6} \b)^2 \times 18.
\end{displaymath}

Adding above three contributions as well as the tree result 1, we get
\begin{equation}
K(\t) = 1+ \frac{i \b^2}{12} (-\frac{\sh\t}{ \ch\t-1}
           -\frac{\sh\t}{2\ch\t-\sqrt{3}}+\frac{\sh\t}{\ch\t}
           +\frac{\sh\t}{\ch\t+1/2}-\frac{\sh\t}{2\ch\t+\sqrt{3}} ) +O(\b^4).
\end{equation}
This satisfies boundary crossing unitarity relation and boundary bootstrap
equation up to $\b^2$ order.
\begin{equation}
K(\t) ~ K(\t-i \pi) = S(2 \t),
  ~~~~~ K(\t)=K(\t+i \pi/3) ~K(\t-i \pi/3) ~ S(2 \t).
\end{equation}

On the other hand, there are four \lq minimal' boundary reflection matrices
which are meromorphic in terms of rapidity variable
are known for $a_2^{(2)}$ model\cite{FK,KCK}.
None of these corresponds to the perturbative result.
A possible exact solution would be the following.
\begin{equation}
K(\t)=[ 1 /2 ] [3/2 ] \sqrt{ \frac{ [1 ]} { [2 ]} }.
\end{equation}
This is one of the \lq  minimal' non-meromorphic solutions
with square root branches which are determined from symmetry principles
such as boundary crossing relation and boundary bootstrap equation.

\section*{V. Discussions}
In this paper, we proposed a method to compute boundary reflection
matrix directly from the two-point correlation function rather
than using the LSZ reduction which is not applicable
to the quantum field theory on a half line.
In our formalism, the unintegrated version of the Neumann Green's function
turns
out to be very useful to extract the asymptotics of the two-point correlation
function in the region far away from the boundary.
This enables us to determine the boundary reflection matrix
for the affine Toda field theory, specifically $a_1^{(1)}$ and $a_2^{(2)}$
models,
with the Neumann boundary condition modulo \lq a mysterious factor half'.

We have also done similar computations for some $a_n^{(1)}$ models
as well as some $d_n^{(1)}$ models.
When the theory has a particle spectrum with more than one mass,
each contribution from three types of diagram in Eq.(\ref{K-III})
has non-meromorphic terms.
According to our partial result, it seems that
for $a_n^{(1)}$ model with $n \geq 3$, the non-meromorphic terms do not add up
to
zero while for $d_n^{(1)}$ theory, they do cancel among themselves exactly and
very nontrivially.

There are many things for future works.
To mention a few, the first thing is to generalize this method
systematically to all loop order, which seems rather straightforward though
the actual evaluation may not be easy.
The second thing is to consider the multi-point correlation functions.
The third thing is to accommodate non-trivial boundary potentials
maintaining the integrability of the model.
Of course, the mysterious factor half and the non-vanishing of the
non-meromorphic
terms for $a_n^{(1)}$ model with $n \geq 3$ need much attention.

\section*{Acknowledgement}
I would like to thank Jihn E Kim and Q-Han Park for encouragement.
I am also grateful to Ed Corrigan and Ryu Sasaki for discussions
and suggestions as well as a critical reading of the original manuscript
and Zheng-Mao Sheng for discussions.
This work was supported by Korea Science and Engineering Foundation
and in part by the University of Durham.

\end{document}